\documentclass[paper]{ieice}
\usepackage[dvips]{graphicx}
\usepackage[fleqn]{amsmath}
\usepackage{newtxtext}
\usepackage[varg]{newtxmath}

\field{D}
\SpecialSection{Parallel, Distributed, and Reconfigurable Computing, and Networking}
\title{RVCoreP : An optimized RISC-V soft processor of five-stage pipelining}
\authorlist{%
  \authorentry[miyazaki@arch.cs.titech.ac.jp]{Hiromu Miyazaki}{s}{titech}\MembershipNumber{1900691}
  \authorentry[kanamori@arch.cs.titech.ac.jp]{Takuto Kanamori}{n}{titech}\MembershipNumber{}
  \authorentry[ashraful@arch.cs.titech.ac.jp]{Md Ashraful Islam}{n}{titech}\MembershipNumber{}
  \authorentry[kise@c.titech.ac.jp]{Kenji Kise}{m}{titech}\MembershipNumber{0305928}
}
\affiliate[titech]{The authors are with the School of Computing,
Tokyo Institute of Technology.}
\received{2020}{1}{7}
\usepackage{comment}
\usepackage{multirow}
\usepackage{xspace}

\usepackage{listings}

\newcommand{\proposal}{RVCoreP\xspace}

\begin{document}
\maketitle

%%%%%%%%%%%%%%%%%%%%%%%%%%%%%%%%%%%%%%%%%%%%%%%%%%%%%%%%%%%%%%%%%%%%%%%%%%%%%%%%%%%%%%%%%%
\begin{summary}
RISC-V is a RISC based open and loyalty free instruction set architecture
which has been developed since 2010,
and can be used for cost-effective soft processors on FPGAs.
The basic 32-bit integer instruction set in RISC-V is defined as RV32I, 
which is sufficient to support the operating system environment
and suits for embedded systems.

In this paper, 
we propose an optimized RV32I soft processor named \proposal 
adopting five-stage pipelining.
The processor applies three effective optimization methods 
to improve the operating frequency.
These methods are 
instruction fetch unit optimization
including pipelined branch prediction mechanism, ALU optimization,
and data alignment and sign-extension optimization for data memory output.
We implement \proposal in Verilog HDL
and verify the behavior using Verilog simulation and an actual Xilinx Atrix-7 FPGA board.
We evaluate IPC (instructions per cycle), operating frequency, 
hardware resource utilization, and processor performance.
From the evaluation results,
we show that \proposal achieves 30.0\% performance improvement
compared with VexRiscv, 
which is a high-performance and open source RV32I processor 
selected from some related works.
\end{summary}

%%%%%%%%%%%%%%%%%%%%%%%%%%%%%%%%%%%%%%%%%%%%%%%%%%%%%%%%%%%%%%%%%%%%%%%%%%%%%%%%%%%%%%%%%%

%%%%%%%%%%%%%%%%%%%%%%%%%%%%%%%%%%%%%%%%%%%%%%%%%%%%%%%%%%%%%%%%%%%%%%%%%%%%%%%%%%%%%%%%%%
\section{Introduction}

RISC-V~\cite{RISCV} is becoming popular
as an open and loyalty free instruction set architecture (ISA)
which has been developed at the University of California, Berkeley since 2010.
It can be used for cost-effective soft processors on FPGAs
like MicroBlaze~\cite{MicroBlaze} and Nios II~\cite{Nios}.

The RISC-V ISA is defined as a basic integer instruction set
and other extended instruction sets, 
and we can support necessary instruction sets by the application requirements~\cite{riscvmanual}.
The basic 32-bit integer instruction set is defined as {\it RV32I}.
Other typical extended instruction sets are defined as
{\it M} for integer multiplication and division instructions, 
{\it F} for single-precision floating-point ones, 
{\it D} for double-precision floating-point ones, and {\it A} for atomic ones.
In addition to these, a 32-bit general-purpose instruction set is defined as {\it RV32G}
as the set of RV32I, M, A, F, and D.
This is an instruction set architecture 
for general-purpose computing systems of a broad range.
{\it RV64G} is a 64-bit version of a general-purpose instruction set.

Among these instruction sets, we focus on RV32I in this paper because 
it is sufficient to support the operating system environment 
and suits for embedded systems.
RV32I can emulate other extensions of M, F, and D, 
and can be configured with fewer hardware resources than processors supporting RV32G.
Although several soft processors 
that support RV32I have been released~\cite{riscveval},
they are not highly optimized for FPGAs.

In this paper,
we propose an optimized RV32I soft processor named {\bf \proposal}
of five-stage pipelining which is highly optimized for FPGAs.
The main contributions of this paper are as follows.
\begin{itemize}
\item
  We propose an optimized RV32I soft processor
  of five-stage pipelining highly optimized for FPGAs.
  To improve the operating frequency, 
  it applies instruction fetch unit optimization including pipelined branch prediction mechanism,
  ALU optimization, and data alignment and sign-extension optimization for data memory output.
\item
  We implement the proposal in Verilog HDL 
  and evaluate IPC (instructions per cycle), operating frequency, 
  hardware resource utilization, and processor performance.
  From the evaluation results,
  we show that the proposed processor achieves much better performance than 
  VexRiscv, which is a high-performance and open source RV32I processor.
\end{itemize}

%%%%%%%%%%%%%%%%%%%%%%%%%%%%%%%%%%%%%%%%%%%%%%%%%%%%%%%%%%%%%%%%%%%%%%%%%%%%%%%%%%%%%%%%%%

\begin{figure*}[tb]
  \begin{center}
    \includegraphics[clip,width=\linewidth]{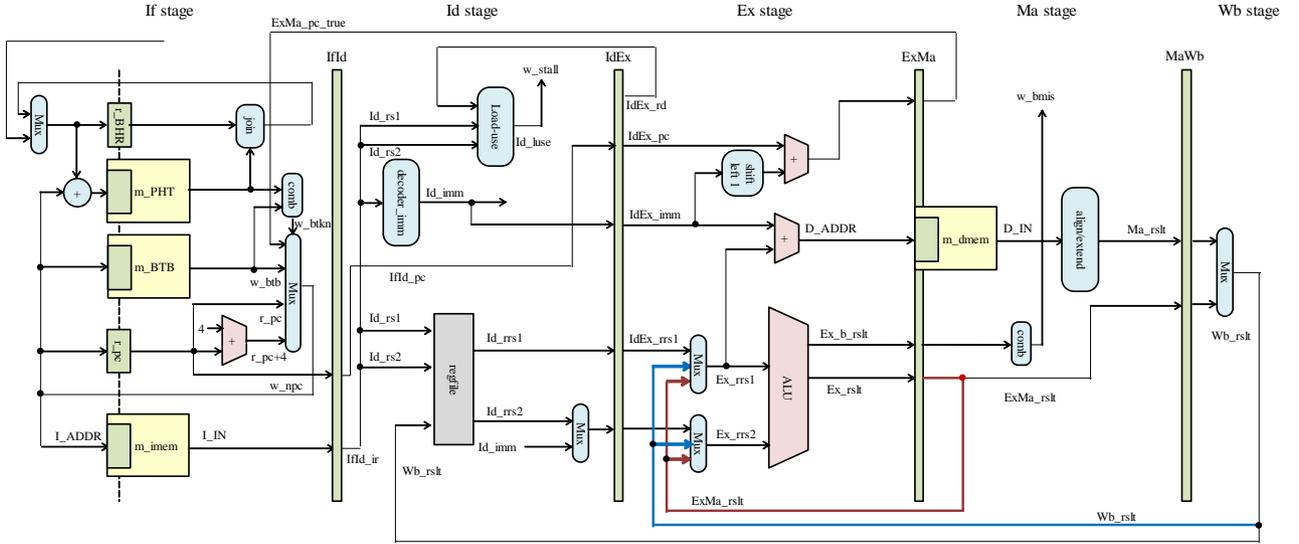}
    \caption{A block diagram of typical five-stage pipelined processor (baseline).}
    \label{fig:baselineproc}
  \end{center}
\end{figure*}

%%%%%%%%%%%%%%%%%%%%%%%%%%%%%%%%%%%%%%%%%%%%%%%%%%%%%%%%%%%%%%%%%%%%%%%%%%%%%%%%%%%%%%%%%%

\section{Related works}

%% \subsection{Rocket Core}

Rocket Core~\cite{rocket} is a RISC-V in-order scalar processor
developed by the University of California, Berkeley.
It is a pipelined processor supporting RV32G and RV64G.
It supports processing of privilege levels,
and has an MMU (memory management unit) with virtual memory and data cache, 
and a branch prediction unit.
Because of this rich functionality and hard customization, 
it is not suitable for embedded systems.

Rocket Core has another drawback.
It is written in Chisel~\cite{chisel}, 
a domain-specific language based on Scala.
Because Chisel is a new hardware description language since 2012,
it may be difficult for hardware developers who have not mastered Chisel 
to change the design effortlessly.
According to the work~\cite{riscveval},
Verilog HDL and SystemVerilog are the dominant languages
used to implement the processors,
and they may be the best choice 
for easy-to-use processor implementations.
Therefore, we implement our processors in Verilog HDL, 
a dominant hardware description language.

%% \subsection{VexRiscv}

VexRiscv~\cite{vexriscv} is a RISC-V pipelined soft processor
supporting RV32I.
The integer multiplication and division, other extensions,
and the MMU with instruction cache and data cache can be added as options.
In addition, the branch prediction scheme,
implementation choice of shift instruction,
data forwarding path, and so on can be tuned for implementation.
VexRiscv is written in an open source and new hardware description language
called SpinalHDL~\cite{spinalhdl},
and the corresponding RTL description can be generated as a Verilog HDL file.
Since the generated Verilog HDL code is not hierarchical,
debugging and understanding this generated code is not easy.

VexRiscv has won the 1st place 
at the highest-performance implementation category of 
the RISC-V SoftCPU Contest in 2018 
hosted by the RISC-V Foundation~\cite{cpucontest}.
Therefore, it is an optimized soft processor for high-performance,
and the highest performance RV32I soft processor available as an open source 
as far as we know.
We use VexRiscv as a reference for making the comparison 
with our proposed processors.

%% \subsection{The other RISC-V processor}

There are other RISC-V processors for education
such as riscv-mini~\cite{riscvmini} and Sodor Processor~\cite{sodor}
both are developed by the University of California, Berkeley,
and Clarvi~\cite{clarvi} developed by the University of Cambridge.
These educational RISC-V processors are easy-to-use,
but their performance is not high as VexRiscv 
because they are not highly optimized for high-performance.

%%%%%%%%%%%%%%%%%%%%%%%%%%%%%%%%%%%%%%%%%%%%%%%%%%%%%%%%%%%%%%%%%%%%%%%%%%%%%%%%%%%%%%%%%
\section{Design of a typical five-stage pipelined processor}

We design a typical five-stage pipelined processor with branch prediction
referring \cite{PataHene_RISCV}, and this design is used as a {\it baseline} 
for the proposal.

Fig.\ref{fig:baselineproc} shows a block diagram of
the baseline consists of five-stage indicated by
the instruction fetch stage (If stage), instruction decode stage (Id stage),
instruction execution stage (Ex stage), memory access stage (Ma stage),
and write back stage (Wb stage).

The green rectangles are registers that are updated at the positive clock edge.
The yellow rectangles are modules including the memory 
which is composed of block RAM on Xilinx FPGA.
The gray rectangle is a register file read asynchronously
consisting of 32 registers.
The red modules are an ALU or adders,
and the other blue modules are combinational circuits.

The baseline has an instruction memory named {\it m\_imem}
shown at the bottom left of the figure,
and a data memory named {\it m\_dmem} shown at the right of the figure.

The branch prediction scheme is gshare~\cite{gshare}
which contains a branch history register (BHR) named {\it r\_BHR},
a pattern history table (PHT) named {\it m\_PHT}, 
and a branch target buffer (BTB) named {\it m\_BTB}.
To mitigate the data hazard, it has two forwarding paths.
The red path from Ma stage to Ex stage provides 
the register value for the next dependent instruction.
Similarly, the blue path provides register value from the Wb stage to the Ex stage.

In the If stage, the instruction is fetched from the instruction memory
using the program counter (PC) as an address.
The register for PC named {\it r\_pc} is updated 
in every cycle with the next PC value named {\it w\_npc}.

There are four candidates for w\_npc in following descending priority order.
The highest priority one is the correct PC value named {\it ExMa\_pc\_true}
from the Ma stage.
The second priority one is the current PC value from r\_pc in case of pipeline stalling.
The third priority one is the branch target address named {\it w\_btb} 
which is output from the BTB.
The lowest priority one is r\_pc+4 for the instruction of the next address.

There are three control signals to select the proper one among four candidates.
The first signal is named {\it w\_bmis} 
which indicates whether a branch misprediction has occurred.
The second one is named {\it w\_stall} for pipeline stalling 
due to the data dependency on the load instruction.
The third one is named {\it w\_btkn} from branch predictor
to provide a prediction result as predicted taken or not taken.

In the baseline, 
the path that determines the next PC value from the four candidates
through a multiplexer using three control signals
is the critical path that determines the maximum operating frequency.
The next critical path is
the data path to store the executed result in the Ex stage from ALU
which uses two data forwarding values.
Another slow path is aligning and sign-extending the values of reading data
from the data memory on the Ma stage which will be stored into 
the MaWb pipeline register.

In our proposed processor,
the operating frequency is improved by optimizing these critical paths.

%%%%%%%%%%%%%%%%%%%%%%%%%%%%%%%%%%%%%%%%%%%%%%%%%%%%%%%%%%%%%%%%%%%%%%%%%%%%%%%%%%%%%%%%%%
\section{Design and implementation of \proposal}

In this section, we propose  an optimized RV32I soft processor named 
{\it \proposal} ({\bf R}ISC-{\bf V} {\bf core} {\bf p}ipelined version).
Firstly, we describe three optimization methods.
Then, we describe the design and implementation of our proposal.

%%%%%%%%%%%%%%%%%%%%%%%%%%%%%%%%%%%%%%%%%%%%%%%%%%%%%%%%%%%%%%%%%%%%%%%%%%%%%%%%%%%%%%%%%%
\subsection{ALU optimization}

The data path to store the executed result in ALU using two data forwarding values 
to the ExMa pipeline register is the critical path in the baseline design.
To mitigate the delay of this critical path, we discuss the ALU optimization scheme.

According to the related work~\cite{alu32},
the circuit speed is faster by using exclusive OR instead of multiplexer
to select the operation result for the ALU optimization on FPGA.
Therefore, in this design of \proposal,
exclusive OR is used to select the 32-bit executed result of ALU.

As mentioned in the related work~\cite{fpgahdl}, 
one-hot encoding is used instead of the usual binary encoding
for the control signal generation to select the ALU calculation result.
As only one bit of the bit vector is 1 and the other bits are 0,
and the control decisions are determined by the corresponding flip-flop bit in parallel.
Therefore, the proposal adopts a one-hot encoding for ALU.

The code \ref{baselinealu.v} is the simplified description of a typical ALU in the baseline
where some operations of RV32I are excluded.
The register named {\it r\_rslt} is the executed result of ALU.
This value is selected by the 3-bit signal named {\it sel},
which is described in a case statement from line 7 to line 16.
Since this description is mapped to hardware as a multiplexer
that selects one from eight values, this circuit takes a certain time through several LUTs.

The code \ref{proposalalu.v} is the simplified description of the optimized ALU
equivalent to the previous description in the code \ref{baselinealu.v}.
The executed result of ALU named {\it rslt} is selected from eight values including 0
using exclusive OR on line 12.
Each selected value is determined in advance using small multiplexers 
by a one-hot encoded selection signal named {\it sel} of 8-bit.
Since this scheme can select a value without using a large multiplexer,
this circuit is faster than the typical one.

The preliminary evaluation of the operating frequency
of the ALU alone targetting Xilinx Artix-7 FPGA showed that 
the frequency of the typical ALU was 230MHz while
the frequency of the optimized ALU was 240MHz.
This optimization is expected to improve the operating frequency of ALU by about 10MHz.

%%%%%%%%%%%%%%%%%%%%%%%%%%%%%%%%%%%%%%%%%%%%%%%%%%%%%%%%%%%%%%%%%%%%%%%%%%%%%%%%%%%%%%%%%%

\begin{figure}[tb]
  \begin{center}
\begin{lstlisting}[label=baselinealu.v,
      caption=The simplified description of a typical ALU.]
module ALU (in1, in2, sel, rslt);
  input  wire [31:0] in1, in2;
  input  wire [2:0]  sel;
  output wire [31:0] rslt;    
  reg [31:0] r_rslt;
  always @(*) begin
    case(sel)
      0 : r_rslt = in1 + in2;       // add
      1 : r_rslt = in1 - in2;       // sub
      2 : r_rslt = in1 ^ in2;       // ex-or
      3 : r_rslt = in1 | in2;       // or
      4 : r_rslt = in1 & in2;       // and
      5 : r_rslt = in1 << in2[4:0]; // shift left
      6 : r_rslt = in1 >> in2[4:0]; // shift right
      default : r_rslt = 0;
    endcase
  end
  assign rslt = r_rslt;
endmodule
\end{lstlisting}
  \end{center}
\end{figure}

%%%%%%%%%%%%%%%%%%%%%%%%%%%%%%%%%%%%%%%%%%%%%%%%%%%%%%%%%%%%%%%%%%%%%%%%%%%%%%%%%%%%%%%%%%

\begin{figure}[tb]
  \begin{center}
\begin{lstlisting}[label=proposalalu.v,
      caption=The simplified description of the optimized ALU.]
module ALU_opt (in1, in2, sel, rslt);
  input  wire [31:0] in1, in2;
  input  wire [7:0]  sel;
  output wire [31:0] rslt;
  wire [31:0] w0 = (sel[0]) ? in1 + in2       : 0;
  wire [31:0] w1 = (sel[1]) ? in1 - in2       : 0;
  wire [31:0] w2 = (sel[2]) ? in1 ^ in2       : 0;
  wire [31:0] w3 = (sel[3]) ? in1 | in2       : 0;
  wire [31:0] w4 = (sel[4]) ? in1 & in2       : 0;
  wire [31:0] w5 = (sel[5]) ? in1 >> in2[4:0] : 0;
  wire [31:0] w6 = (sel[6]) ? in1 << in2[4:0] : 0;
  assign rslt = w0 ^ w1 ^ w2 ^ w3 ^ w4 ^ w5 ^ w6;
endmodule
\end{lstlisting}
  \end{center}
\end{figure}

%%%%%%%%%%%%%%%%%%%%%%%%%%%%%%%%%%%%%%%%%%%%%%%%%%%%%%%%%%%%%%%%%%%%%%%%%%%%%%%%%%%%%%%%%%
\subsection{Alignment and sign-extension optimization}

After applying the ALU optimization 
and the instruction fetch unit optimization described later,
the critical path is the data memory access and the alignment and sign-extension 
of the reading data which is computed using the combinational circuit named {\it align/extend}
on the Ma stage in Fig.\ref{fig:baselineproc}.

RV32I has five load instructions which are
load byte (LB) to load 8-bit signed data,
load byte unsigned (LBU) to load 8-bit unsigned data,
load halfword (LH) to load 16-bit signed data,
load halfword unsigned (LHU) to load 16-bit unsigned data,
and load word (LW) to load 32-bit data.
Therefore, align/extend unit has to align the loaded data by shifting 8, 16, or 24 bits right
depends on the memory address and operation code of the load instructions.
Then, sign-extension or zero extension is needed for load byte and load halfword instructions.
Finally, the unit selects a proper value using a large multiplexer depends on the operation 
code of the instruction.

We optimized the alignment and sign-extension using 
the similar approach to the ALU optimization 
which is one-hot encoding and using exclusive OR for value selection.

%%%%%%%%%%%%%%%%%%%%%%%%%%%%%%%%%%%%%%%%%%%%%%%%%%%%%%%%%%%%%%%%%%%%%%%%%%%%%%%%%%%%%%%%%%
\subsection{Instruction fetch unit optimization}

We propose the two-stage pipelining of the branch predictor
to improve the operating frequency of the instruction fetch unit, which 
contains the critical path on the baseline processor.

The related works~\cite{pipelinebpred,matsuibpred} have shown that
the pipelining of the branch predictor can improve the operating frequency 
of the soft processor when a complex branch predictor is used.
The similar approach is applied to the proposed branch prediction
including gshare and BTB.

%%%%%%%%%%%%%%%%%%%%%%%%%%%%%%%%%%%%%%%%%%%%%%%%%%%%%%%%%%%%%%%%%%%%%%%%%%%%%%%%%%%%%%%%%%

\begin{figure}[tb]
  \begin{center}
    \includegraphics[clip,width=\linewidth]{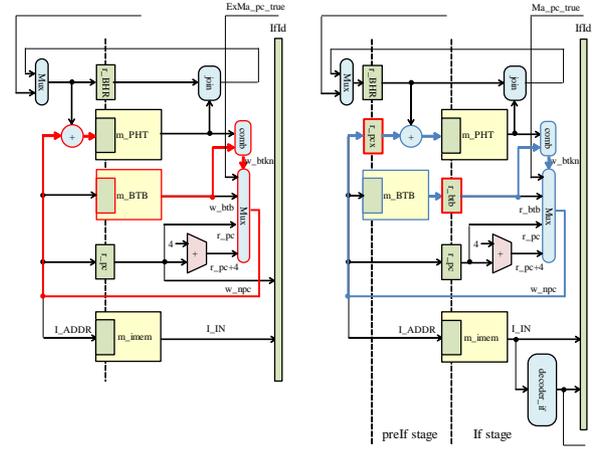}
    \caption{A general configuration and the two-stage pipelining one
for branch prediction mechanism including gshare, BTB and instruction memory.}
    \label{fig:gshare}
  \end{center}
\end{figure}

Fig.\ref{fig:gshare} (a) shows a block diagram 
of a general branch predictor and BTB in the baseline where the prediction is made in
a single cycle.
In gshare branch predictor,
the index to access the PHT named {\it m\_PHT} is 
obtained by exclusive OR of PC and BHR named {\it r\_BHR}.
If the fetched instruction is predicted as a conditional branch instruction using the BTB,
it updates the BHR speculatively using the branch prediction in the If stage. 

The combinational circuit named {\it join} shifts BHR left by 1-bit,
and connects the branch prediction result to the least significant bit of the BHR.
If the branch prediction missed,
the BHR is updated with the correct branch history.
The combinational circuit named {\it comb} that receives the value read from the BTB
and the value read from the PHT generates the branch prediction 
named {\it w\_btkn}.

The critical path of a general branch prediction mechanism
is the red data path in Fig.\ref{fig:gshare} (a)
which includes the access of the BTB, three LUTs, a multiplexer, wiring delays, and clock skew.
In our preliminary evaluation,
the access of the BTB composed of block RAM 
on a Xilinx Artix-7 FPGA takes about 2.54ns on the red path.
Since the access of one LUT takes about 0.12ns, 
the access to the three LUTs necessary on the red path takes about 0.36ns.
Also, the access to a multiplexer implemented using hard macro takes about 0.24ns,
and other wiring delays and clock skew takes about 3ns.
Therefore, the total delay of the path exceeds 6.1ns (165MHz) by adding the above delays.

To improve the operating frequency for the proposed processor,
we split this critical path by two registers.

%%%%%%%%%%%%%%%%%%%%%%%%%%%%%%%%%%%%%%%%%%%%%%%%%%%%%%%%%%%%%%%%%%%%%%%%%%%%%%%%%%%%%%%%%%

%%%%%%%%%%%%%%%%%%%%%%%%%%%%%%%%%%%%%%%%%%%%%%%%%%%%%%%%%%%%%%%%%%%%%%%%%%%%%%%%%%%%%%%%%%
\begin{figure}[tb]
  \begin{center}
    \includegraphics[clip,width=\linewidth]{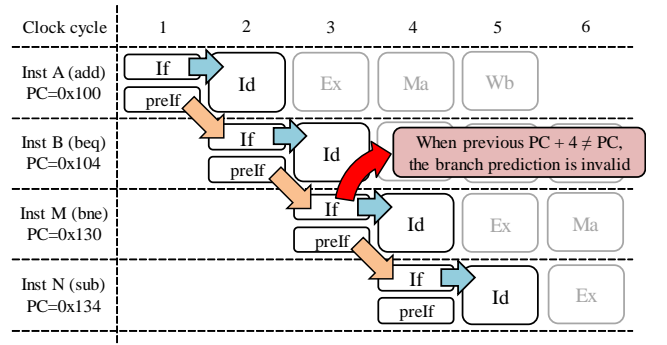}
    \caption{The pipeline diagram of instruction fetching
using pipelined branch prediction mechanism in \proposal.}
    \label{fig:fetchchart}
  \end{center}
\end{figure}

%%%%%%%%%%%%%%%%%%%%%%%%%%%%%%%%%%%%%%%%%%%%%%%%%%%%%%%%%%%%%%%%%%%%%%%%%%%%%%%%%%%%%%%%%%

Fig.\ref{fig:gshare} (b) shows the block diagram 
of the pipelined gshare and pipelined BTB for \proposal.
The red critical path in Fig.\ref{fig:gshare} (a) is divided into three paths 
by two inserted registers named {\it r\_btb} and {\it r\_pcx}.
The data acquired from the BTB is stored in the register r\_btb,
and the register r\_pcx is inserted before exclusive OR 
to generate the index of PHT.

It takes two cycles to determine the value of the next PC in the instruction fetch stage.
In the first cycle in the {\it preIf} stage,
accessing the BTB and exclusive OR processing to determine the PHT index are performed.
In the second cycle in the If stage,
the value of the next PC is determined by using the results from the preIf stage
and the instruction is fetched from the instruction memory.

%%%%%%%%%%%%%%%%%%%%%%%%%%%%%%%%%%%%%%%%%%%%%%%%%%%%%%%%%%%%%%%%%%%%%%%%%%%%%%%%%%%%%%%%%%
\begin{figure*}[tb]
  \begin{center}
    \includegraphics[clip,width=\linewidth]{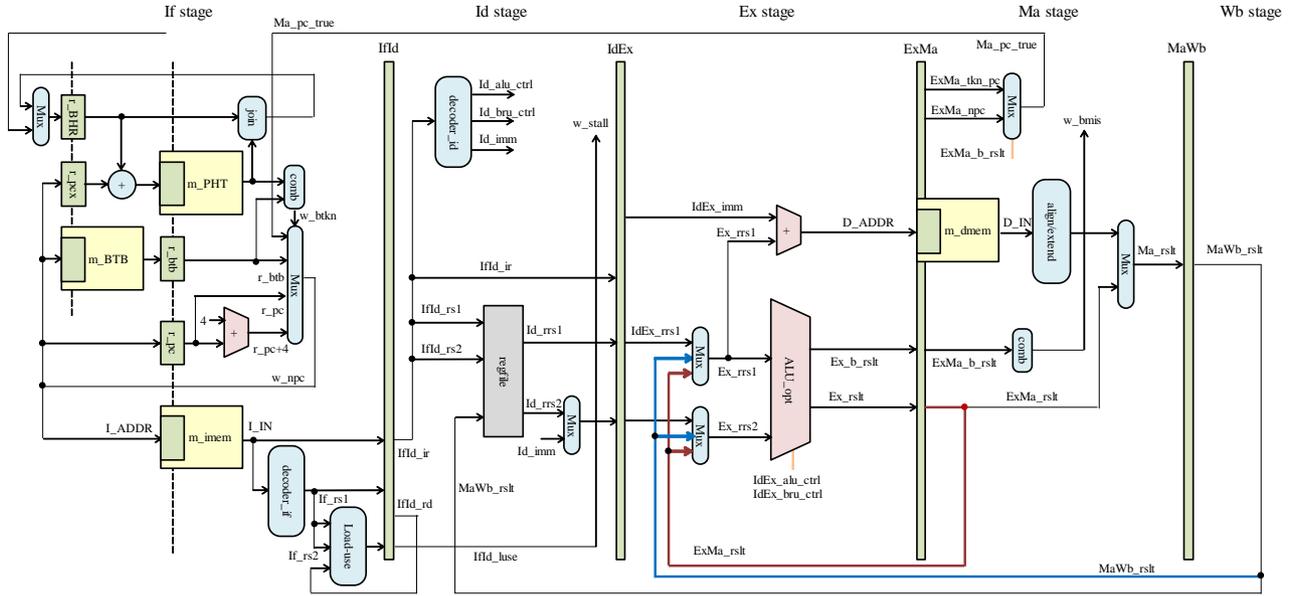}
    \caption{The block diagram of the proposed processor named \proposal.}
    \label{fig:proposalproc}
  \end{center}
\end{figure*}

%%%%%%%%%%%%%%%%%%%%%%%%%%%%%%%%%%%%%%%%%%%%%%%%%%%%%%%%%%%%%%%%%%%%%%%%%%%%%%%%%%%%%%%%%%

Fig.\ref{fig:fetchchart} shows the pipeline diagram of instruction fetching
using the pipelined branch prediction in \proposal.
The rectangles written as preIf represents the processing of the preIf stage,
and the rectangles written as If represents the processing of the If stage.
Assuming that four instructions are fetched 
in the order of {\it Inst A}, {\it Inst B}, {\it Inst M}, and {\it Inst N}.
Inst A and Inst N are add and sub instructions,
and these instruction addresses are 0x100 and 0x134, respectively.
Inst B and Inst M are beq (branch if equal) and 
bne (branch if not equal) instruction,
and these instruction addresses are 0x104 and 0x130, respectively.
The next PC of Inst B is 0x130 when the branch is taken.

In the clock cycle 1 when the value of PC is 0x100,
the If stage for Inst A and the preIf stage for the next instruction are executed.
In the preIf stage,
the value of BTB and PHT index used in the If stage for the next instruction 0x104
are prepared by using the current PC value of 0x100.

In the clock cycle 2 when the value of PC is 0x104,
the If stage for Inst B and the preIf stage for the next instruction are executed.
In the If stage for Inst B,
the next PC value is determined 
by using the value prepared in the preIf stage one cycle before.
In the preIf stage,
the value of BTB and PHT index used in the If stage for the next instruction 0x108
are prepared by using the current PC value of 0x104.
Since Inst B is a conditional branch and assuming that it is predicted as taken,
the next PC value is 0x130.

In the clock cycle 3 when the value of PC is 0x130,
the If stage for Inst M and the preIf stage for the next instruction are executed.
In the If stage for Inst M,
the next PC value is determined as well but branch prediction is invalid,
because the value prepared in the preIf stage one cycle before
is for the instruction whose address is 0x108,
and this value cannot be used in the If stage for Inst M whose address is 0x130.
Therefore, if the value obtained by adding 4 to the previous PC
does not match the current PC value, the branch prediction is invalid.

From the above,
the PC value used for BTB access is the one cycle earlier value of PC,
and the PC value used for PHT access is the value one cycle before
the branch prediction is output.
As a result, gshare outputs a prediction in 2 cycles.
The BTB entry is updated using a value obtained
by subtracting 4 from the PC value of the branch instruction.
When updating a PHT entry,
we have to keep the PHT index value used for the prediction 
and to update the PHT entry using this index
when the actual branch outcome will be available.

The prediction accuracy might drop slightly
due to the adverse effect of this optimization
to make a prediction and update the index 
with the one cycle earlier value of PC.

%%%%%%%%%%%%%%%%%%%%%%%%%%%%%%%%%%%%%%%%%%%%%%%%%%%%%%%%%%%%%%%%%%%%%%%%%%%%%%%%%%%%%%%%%%
\subsection{\proposal soft processor}

Fig.\ref{fig:proposalproc} shows the block diagram of \proposal which
is a five-stage pipelined processor
including an instruction memory, a data memory, pipelined gshare 
and pipelined BTB.
The ALU optimization,
the alignment and sign-extension optimization, and
the instruction fetch unit optimization are applied to the proposal.
The unit named {\it ALU\_opt} in Fig.\ref{fig:proposalproc} is the optimized ALU.

The detection timing of the load-use dependency between a load instruction and 
the following instruction using the loaded data
is changed from the Id state on the baseline in Fig.\ref{fig:baselineproc}
to the If stage using the combinational circuit named {\it Load-use}.
To support the detection, a part of instruction decoder named {\it decoder\_if} is
implemented in the If stage.
decoder \_if decodes two source registers and one destination register for one instruction,
and generates the write signals for the register file and data memory.
This partial decoding of instruction in If stage allows us
to detect the data dependency including load-use dependency in advance.

%%%%%%%%%%%%%%%%%%%%%%%%%%%%%%%%%%%%%%%%%%%%%%%%%%%%%%%%%%%%%%%%%%%%%%%%%%%%%%%%%%%%%%%%%%
\begin{figure}[tb]
  \begin{center}
    \includegraphics[clip,width=\linewidth]{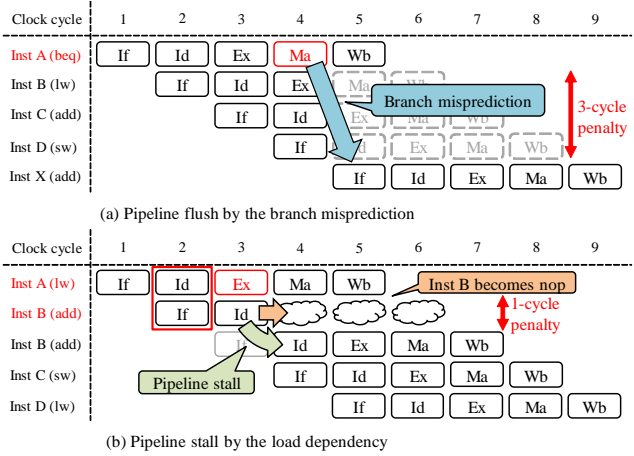}
    \caption{The pipeline diagrams of the proposed processor.}
    \label{fig:pipelinechart}
  \end{center}
\end{figure}

%%%%%%%%%%%%%%%%%%%%%%%%%%%%%%%%%%%%%%%%%%%%%%%%%%%%%%%%%%%%%%%%%%%%%%%%%%%%%%%%%%%%%%%%%%

\begin{table*}[tb]
  \begin{center}
    \caption{The evaluation results of IPC and branch prediction accuracy obtained by Verilog simulation.}
    \begin{tabular}{l|c|c|c|c|c|c|c|c|c} \hline
      \multirow{2}{*}{Label} & \multicolumn{4}{|c|}{Dhrystone} & \multicolumn{4}{|c|}{Coremark} & \multirow{2}{*}{Average IPC} \\ \cline{2-9}
      & \multicolumn{1}{|c|}{IPC} & prediction hit & prediction miss & hit rate
      & \multicolumn{1}{|c|}{IPC} & prediction hit & prediction miss & hit rate & \\ \hline
      VR-nobp    & 0.661 & N/A     & N/A    & N/A   & 0.591 & N/A     & N/A     & N/A   & 0.626 \\ \hline
      VR-bp      & 0.836 & 146,180 & 29,452 & 0.832 & 0.766 & 348,010 & 109,701 & 0.760 & 0.801 \\ \hline
      RVP-simple & 0.946 & 205,127 & 12,507 & 0.943 & 0.828 & 366,726 &  91,247 & 0.801 & 0.887 \\ \hline
      RVP-optALU & 0.946 & 205,127 & 12,507 & 0.943 & 0.828 & 366,726 &  91,247 & 0.801 & 0.887 \\ \hline
      RVP-optIF  & 0.935 & 201,153 & 16,481 & 0.924 & 0.823 & 363,439 &  94,534 & 0.794 & 0.879 \\ \hline
      RVP-optALL & 0.935 & 201,153 & 16,481 & 0.924 & 0.823 & 363,439 &  94,534 & 0.794 & 0.879 \\ \hline
    \end{tabular}
    \label{tab:ipcbpresult}
  \end{center}
\end{table*}

%%%%%%%%%%%%%%%%%%%%%%%%%%%%%%%%%%%%%%%%%%%%%%%%%%%%%%%%%%%%%%%%%%%%%%%%%%%%%%%%%%%%%%%%%%

Fig.\ref{fig:pipelinechart} shows the pipeline diagrams of the proposed processor.
Fig.\ref{fig:pipelinechart} (a) shows the case where 
the pipeline is flushed due to a branch prediction miss.
In the branch prediction mechanism,
the branch target address from the BTB is used
when the BTB is hit and the branch is predicted to be taken.
The correct branch destination address calculation 
and check whether the branch prediction is correct or not
is executed in the Ex stage and stored in the ExMa pipeline register.
If the branch instruction is at the Ma stage and the branch prediction missed,
the instructions in the If stage, Id stage, and Ex stage are flushed,
which incurs a 3-cycle penalty.

Fig.\ref{fig:pipelinechart} (b) shows the case where 
the pipeline stalls due to the load-use dependency.
In that case, the dependency is avoided by 
stalling the instruction following the load instruction.
Using the decoder\_if to partially decode the instruction in the If stage
helps to detect the dependency by load instruction in Id stage 
and an instruction in the If stage,
and the detection result is stored in the IdEx pipeline register.
If the load instruction is in the Ex stage 
and there is a data dependency on load instruction,
this processor inserts a bubble in IdEx pipeline register, 
and stall instructions in the If stage and Id stage,
which incurs a one-cycle penalty.

\section{Verification and evaluation}

\subsection{Verification}

We verified the implemented RTL code by Verilog simulation.
A RISC-V processor simulator modeling a conservative multi-cycle processor 
named {\it SimRV} that we implemented in C++ is used as the reference model.

SimRV outputs the PC value, the executed instruction,
and the 32 values stored in the register file, when a RISC-V program binary is given.
By executing the same binary using SimRV and Verilog simulation for our designed processors,
log files of the same format can be output.
We executed the two benchmark binaries used in the evaluation described later,
and compared each log file.
We have confirmed that 
their values in two log files match and the programs are executing correctly.

In addition to the verification through simulations, we verified the
behavior of the designed processor using an FPGA board.
The same RISC-V program binary used for Verilog simulation is executed 
on the actual Xilinx Atrix-7 FPGA board, 
and we have confirmed that
the ASCII character output of the execution results via a serial communication 
had matched to the correct result,
and confirmed that the numbers of execution cycles and executed instructions 
are also matched.

%%%%%%%%%%%%%%%%%%%%%%%%%%%%%%%%%%%%%%%%%%%%%%%%%%%%%%%%%%%%%%%%%%%%%%%%%%%%%%%%%%%%%%%%%%

\begin{table*}[tb]
  \begin{center}
    \caption{The evaluation results of frequency, hardware resource utilization,
      and performance.}
    \begin{tabular}{l|c|c|c|c|c|c|c|c} \hline
      Label & \begin{tabular}{l} Operating \\ frequency    \end{tabular}
            & \begin{tabular}{l} Slice \\ LUT              \end{tabular}
            & \begin{tabular}{l} Slice \\ register         \end{tabular}
            & \begin{tabular}{l} Slice                     \end{tabular}
            & \begin{tabular}{l} Increase rate \\ of slice \end{tabular}
            & \begin{tabular}{l} Average \\ IPC            \end{tabular}
            & \begin{tabular}{l} Processor \\ performance  \end{tabular}
            & \begin{tabular}{l} Normalized \\ performance \end{tabular} \\ \hline
      VR-nobp    & 205 & 936   & 562 & 284 & 1.000 & 0.626 & 128.4 & 1.000 \\ \hline
      VR-bp      & 140 & 944   & 611 & 300 & 1.056 & 0.801 & 112.1 & 0.873 \\ \hline
      RVP-simple & 160 & 1,020 & 715 & 349 & 1.229 & 0.887 & 141.9 & 1.105 \\ \hline
      RVP-optALU & 170 & 1,070 & 730 & 375 & 1.320 & 0.887 & 150.8 & 1.174 \\ \hline
      RVP-optIF  & 180 & 1,044 & 749 & 390 & 1.373 & 0.879 & 158.2 & 1.232 \\ \hline
      RVP-optALL & 190 & 1,073 & 764 & 397 & 1.398 & 0.879 & 167.0 & 1.300 \\ \hline
    \end{tabular}
    \label{tab:freqhardresult}
  \end{center}
\end{table*}

%%%%%%%%%%%%%%%%%%%%%%%%%%%%%%%%%%%%%%%%%%%%%%%%%%%%%%%%%%%%%%%%%%%%%%%%%%%%%%%%%%%%%%%%%%

\subsection{Evaluation environment}

We implement four versions of the proposed processor in Verilog HDL
and evaluate them in terms of IPC, operating frequency, 
hardware resource utilization, and processor performance.
We also make two configurations for VexRiscv processor that supports RV32I,
and use these configurations for comparative evaluation with the proposals.
The code used for \proposal is Ver.0.4.5.
The code version of VexRiscv processor used for the evaluation is 
{\it SpinalHDL/VexRiscv@ca228a3} 
committed on 26 September 2019 in GitHub page~\cite{vexriscv}.

The four versions of \proposal are named as follows.
The version that applies all the optimizations described above 
is called {\bf RVP-optALL},
and the simple version that does not apply any optimizations
is defined as {\bf RVP-simple}.
The version that applies only the ALU optimization 
and the alignment and sign-extension optimization 
is specified as {\bf RVP-optALU},
and the version that applies only the instruction fetch unit optimization
is defined as {\bf RVP-optIF}.

For the VexRiscv, 
{\bf VR-nobp} denotes the configuration 
without the branch prediction,
and {\bf VR-bp} denotes the configuration 
with the branch prediction.
We set the parameters of VexRiscv as follows
to make the configuration as close as possible to \proposal.
They are reading the register file asynchronously,
using shift instruction 
implemented with a full barrel shifter that performs in one cycle,
and utilizing the data forwarding path.

\proposal has a branch prediction mechanism including
a gshare with a PHT of 8,192 entries and a BTB of 512 entries.
If the branch prediction is enabled in the VexRiscv,
{\it Prediction DYNAMIC\_TARGET} of the option in BranchPlugin is used,
and the number of entries in the direct mapped prediction cache is set to 
512 entries using {\it historyRamSizeLog2} parameter.

%% \subsection{Evaluated System}

To execute the RISC-V program with \proposal,
we create a system including the proposed processor.
This system includes the proposed processor \proposal as shown in Fig.\ref{fig:proposalproc},
an instruction memory, a data memory, 
and the modules for RS-232C serial communication
with a communication buffer.
This system reads the RISC-V program binary and operates for Verilog simulation.
Also, the same program for Verilog simulation runs
on Nexys 4 DDR board with Xilinx Artix-7 FPGA~\cite{nexys4ddr}
that receives the same binary through the serial communication module.
This system can output the same characters as the simulation by serial communication.
The number of lines of code for this system is 1,487,
of which the processor \proposal has 832 lines of codes.
This system is used to evaluate IPC, operating frequency, and hardware resource utilization.
By replacing the VexRiscv processor with the processor part of this system,
VexRiscv is evaluated in the same environment.

%%%%%%%%%%%%%%%%%%%%%%%%%%%%%%%%%%%%%%%%%%%%%%%%%%%%%%%%%%%%%%%%%%%%%%%%%%%%%%%%%%%%%%%%%%

\begin{figure}[tb]
  \begin{center}
    \includegraphics[clip,width=\linewidth]{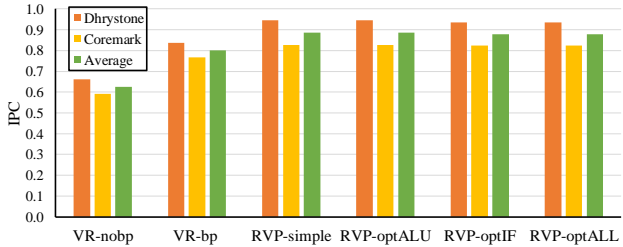}
    \caption{The IPC for each configuration obtained by Verilog simulation.}
    \label{fig:ipcgraph}
  \end{center}
\end{figure}

%%%%%%%%%%%%%%%%%%%%%%%%%%%%%%%%%%%%%%%%%%%%%%%%%%%%%%%%%%%%%%%%%%%%%%%%%%%%%%%%%%%%%%%%%%

%% \subsection{IPC}

IPC is evaluated by Verilog simulation using 
Dhrystone~\cite{Dhrystone} and Coremark~\cite{coremark} as benchmarks.
We used the Dhrystone source code published in riscv-tests~\cite{riscvtests}
and NUMBER\_OF\_RUNS was set to 2000.
The number of executed instructions for Dhrystone is 909,443.
We used the Coremark source code~\cite{riscvcoremark} released for RISC-V
and ITERATIONS was set to 2.
The number of executed instructions for Coremark is 1,481,298.
The source codes of each benchmark are compiled by using the RISC-V RV32I cross compiler.
The RISC-V gcc cross compiler version 8.3.0 has been used, 
and the used optimization flag was -O2.
For benchmark program simulation,
the size of both instruction memory and data memory was set as 32KB.

%% \subsection{Operating frequency and hardware resource utilization}

The operating frequency and hardware resource utilization are evaluated 
targetting Nexys 4 DDR board~\cite{nexys4ddr} having xc7a100tcsg324-1 FPGA 
which is a family of Xilinx Artix-7 FPGA.
Xilinx Vivado 2017.2 is used to evaluate the operating frequency and hardware resource utilization.
Flow\_PerfOptimaized\_high strategy is used for logic synthesis,
and Performance\_ExplorePostRoutePhysOpt strategy is used for placement and routing.
We performed the logic synthesis and placement and routing
by incrementally changing the clock cycle constraint in 5MHz.
The highest frequency that satisfies the constraints is used as the operating frequency 
of the processor.
For hardware resource evaluation,
we used the result of placement and routing at the maximum operating frequency.
The size of instruction memory and data memory is fixed at 4KB for evaluation using FPGA.
To stabilize the operating frequency of the evaluated system,
the placement and routing are performed using only one clock region of the FPGA.

%% \subsection{Processor performance}

The processor performance is calculated by multiplying the average IPC 
by the operating frequency.

%%%%%%%%%%%%%%%%%%%%%%%%%%%%%%%%%%%%%%%%%%%%%%%%%%%%%%%%%%%%%%%%%%%%%%%%%%%%%%%%%%%%%%%%%%

\begin{figure}[tb]
  \begin{center}
    \includegraphics[clip,width=\linewidth]{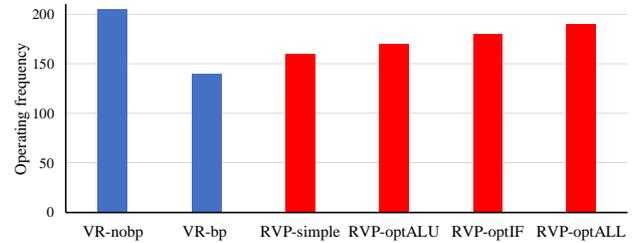}
    \caption{The maximum operating frequency for each configuration on Artix-7 FPGA.}
    \label{fig:freqgraph}
  \end{center}
\end{figure}

%%%%%%%%%%%%%%%%%%%%%%%%%%%%%%%%%%%%%%%%%%%%%%%%%%%%%%%%%%%%%%%%%%%%%%%%%%%%%%%%%%%%%%%%%%
\subsection{Evaluation results}

Table \ref{tab:ipcbpresult} shows the evaluation results of IPC and branch accuracy 
obtained by Verilog simulation.
This shows IPC and the number of prediction hit and miss, 
and prediction hit rate for each of the two benchmarks,
and the average IPC of these two benchmarks.

Regarding IPC and the branch prediction hit rate of each benchmark and average IPC,
the four versions of \proposal outperform the two versions of VexRiscv.
Note that the prediction accuracy of the branch predictor drops 
due to the pipelined branch prediction.
Therefore, the IPC of RVP-simple and RVP-optALU 
is higher than the IPC of RVP-optIF and RVP-optALL.

Fig.\ref{fig:ipcgraph} shows the IPC for each configuration obtained by Verilog simulation.
The orange bars are used for Dhrystone,
the yellow bars for Coremark, and the green bars for the average.
As a whole, Dhrystone has a simpler program structure than Coremark
and a higher branch prediction hit rate.
Therefore, Dhrystone tends to have a higher value of IPC than Coremark.
From this figure, we confirm that
the four versions of \proposal outperform the two versions of VexRiscv.

Table \ref{tab:freqhardresult} summarises the evaluation results of operating frequency, 
hardware resource utilization, and processor performance for Artix-7 FPGA. 
As for slice usage shown in the 5th column, 
VexRiscv is more resource-saving than \proposal as a whole.
The increase rate of slice shown in the 6th column is normalized with VR-nobp as 1.
The slice usage of RVP-optALL is 397,
which is a 39.8\% increase compared to VR-nobp.
Since this slice usage is 2.50\% of the total available slices (which is 15,850) 
on used Artix-7 FPGA,
this increase in slice usage is negligible.
48 LUTs are used as memory (LUTRAM), which are inferred for the register file 
of the processor.
Except for VR-nobp which does not have a branch prediction,
only one block RAM for the tables in branch prediction is used.
In all configurations, the instruction memory and data memory of 4KB 
consist of two block RAMs.

%%%%%%%%%%%%%%%%%%%%%%%%%%%%%%%%%%%%%%%%%%%%%%%%%%%%%%%%%%%%%%%%%%%%%%%%%%%%%%%%%%%%%%%%%%

\begin{figure}[tb]
  \begin{center}
    \includegraphics[clip,width=\linewidth]{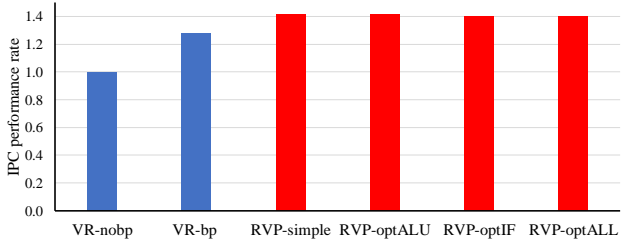}
    \caption{The processor performance by IPC assuming that the operating frequency is the same
      where VR-nobp is normalized as 1.}
    \label{fig:ipcperfgraph}
  \end{center}
\end{figure}

%%%%%%%%%%%%%%%%%%%%%%%%%%%%%%%%%%%%%%%%%%%%%%%%%%%%%%%%%%%%%%%%%%%%%%%%%%%%%%%%%%%%%%%%%%

\begin{figure}[tb]
  \begin{center}
    \includegraphics[clip,width=\linewidth]{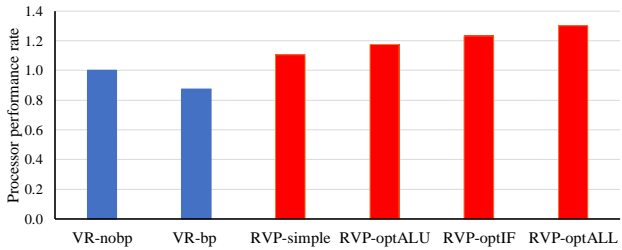}
    \caption{The processor performance on Artix-7 FPGA
      where VR-nobp is normalized as 1.}
    \label{fig:perfgraph}
  \end{center}
\end{figure}

%%%%%%%%%%%%%%%%%%%%%%%%%%%%%%%%%%%%%%%%%%%%%%%%%%%%%%%%%%%%%%%%%%%%%%%%%%%%%%%%%%%%%%%%%%

Fig.\ref{fig:freqgraph} shows the maximum operating frequency 
for each configuration on Artix-7 FPGA.
The configuration of VR-nobp has the highest operating frequency of 205MHz.
It can be seen that
the operating frequency of the four configurations of \proposal
is improved by applying each optimization.
The best frequency of \proposal is 190MHz on RVP-optALL.
Note that among configurations with branch predictions, RVP-optALL achieves much better
operating frequency than VP-bp running at 140MHz.

Fig.\ref{fig:ipcperfgraph} shows the processor performance by IPC
assuming that the operating frequency is the same where VR-nobp is normalized as 1.
RVP-simple and RVP-optALU have the highest performance.
RVP-optALL achieves 40.3\% performance improvement compared to VR-nobp
because VR-nobp does not have a branch prediction and has low IPC.
RVP-optALL achieves 9.74\% performance improvement compared to VR-bp.
The other configurations of \proposal achieve almost the same IPC performance.

Fig.\ref{fig:perfgraph} shows the processor performance on Artix-7 FPGA
where VR-nobp is normalized as 1.
This processor performance considers the operating frequency, 
and each value in the graph is the performance improvement rate from VR-nobp.
RVP-optALL achieves 30.0\% performance improvement compared to VR-nobp,
which is the highest performance configuration of VexRiscv.
The other configurations of \proposal achieve performance improvement of 10\% or more
compared to VR-nobp.

The performance improvement from RVP-simple to RVP-optALL is 1.176.
Therefore, we achieve 17.6\% performance improvement
by using three proposed optimizations.

%%%%%%%%%%%%%%%%%%%%%%%%%%%%%%%%%%%%%%%%%%%%%%%%%%%%%%%%%%%%%%%%%%%%%%%%%%%%%%%%%%%%%%%%%%
\section{Conclusion}

We propose a RISC-V soft processor adopting five-stage pipelining highly optimized for FPGAs.
In the proposed processor,
the instruction fetch unit optimization,
the ALU optimization, and the alignment and sign-extension optimization are
applied as effective methods to improve the operating frequency.
We implement this proposed processor in Verilog HDL
and evaluate IPC, operating frequency, hardware resource utilization, and processor performance
compared with the VexRiscv processor.

From the evaluation results,
the proposed processor RVP-optALL that applied all optimizations
achieves 30.0\% performance improvement
as processor performance considering operating frequency compared with VR-nobp,
which is the highest performance configuration of VexRiscv.
In addition, the proposed optimization method achieves
17.6\% performance improvement in \proposal.

%%%%%%%%%%%%%%%%%%%%%%%%%%%%%%%%%%%%%%%%%%%%%%%%%%%%%%%%%%%%%%%%%%%%%%%%%%%%%%%%%%%%%%%%%%

\section*{Acknowledgments}

This work is supported by JSPS KAKENHI Grant Number JP16H02794.

%%%%%%%%%%%%%%%%%%%%%%%%%%%%%%%%%%%%%%%%%%%%%%%%%%%%%%%%%%%%%%%%%%%%%%%%%%%%%%%%%%%%%%%%%%

\bibliographystyle{ieicetr}% bib style
\bibliography{reference}% your bib database

\begin{thebibliography}{10}

\bibitem{RISCV}
{RISC-V Foundation}, ``{RISC-V | Instruction Set Architecture (ISA)}.''
  https://riscv.org/.

\bibitem{MicroBlaze}
Xilinx, {MicroBlaze Processor Reference Guide}, v2018.2~ed., June\ 2018.

\bibitem{Nios}
Intel, {Nios II Processor Reference Guide}, April\ 2018.

\bibitem{riscvmanual}
A.~Waterman, Y.~Lee, D.A. Patterson, and K.~Asanović, ``{The RISC-V
  Instruction Set Manual, Volume I: User-Level ISA, Version 2.1},'' Tech. Rep.
  UCB/EECS-2016-118, EECS Department, University of California, Berkeley, May\
  2016.

\bibitem{riscveval}
R.~{Höller}, D.~{Haselberger}, D.~{Ballek}, {\em et~al.}, ``{Open-Source
  RISC-V Processor IP Cores for FPGAs — Overview and Evaluation},'' 2019 8th
  Mediterranean Conference on Embedded Computing (MECO), pp.1--6, June\ 2019.

\bibitem{rocket}
K.~Asanović, R.~Avizienis, J.~Bachrach, {\em et~al.}, ``{The Rocket Chip
  Generator},'' Tech. Rep. UCB/EECS-2016-17, EECS Department, University of
  California, Berkeley, Apr\ 2016.

\bibitem{chisel}
J.~{Bachrach}, H.~{Vo}, B.~{Richards}, {\em et~al.}, ``{Chisel: Constructing
  hardware in a Scala embedded language},'' DAC Design Automation Conference
  2012, pp.1212--1221, June\ 2012.

\bibitem{vexriscv}
{SpinalHDL}, ``{VexRiscv: A FPGA friendly 32 bit RISC-V CPU implementation}.''
  https://github.com/SpinalHDL/VexRiscv.

\bibitem{spinalhdl}
{SpinalHDL}, ``{SpinalHDL: An open source high-level hardware description
  language}.'' https://github.com/SpinalHDL/SpinalHDL.

\bibitem{cpucontest}
{RISC-V Foundation}, ``{RISC-V SoftCPU Contest, October 8, 2018}.''
  https://riscv.org/2018/10/risc-v-contest/.

\bibitem{riscvmini}
{University of California, Berkeley}, ``{riscv-mini: Simple RISC-V 3-stage
  Pipeline in Chisel}.'' https://github.com/ucb-bar/riscv-mini.

\bibitem{sodor}
{University of California, Berkeley}, ``{The Sodor Processor: educational
  microarchitectures for risc-v isa}.'' https://github.com/ucb-bar/riscv-sodor.

\bibitem{clarvi}
{University of Cambridge}, ``{Clarvi: simple RISC-V processor for teaching}.''
  https://github.com/ucam-comparch/clarvi.

\bibitem{PataHene_RISCV}
D.A. Patterson and J.L. Hennessy, {Computer Organization and Design The
  Hardware / Software Interface, RISC-V Edition}, Morgan Kaufmann, 2018.

\bibitem{gshare}
S.~McFarling, ``{Combining branch predictors},'' tech. rep., Technical Report
  TN-36, Digital Western Research Laboratory, 1993.

\bibitem{alu32}
P.~Metzgen, ``{A High Performance 32-bit ALU for Programmable Logic},''
  Proceedings of the 2004 ACM/SIGDA 12th International Symposium on Field
  Programmable Gate Arrays, FPGA '04, New York, NY, USA, pp.61--70, ACM, 2004.

\bibitem{fpgahdl}
Xilinx, {HDL Synthesis for FPGAs Design Guide}, 1995.

\bibitem{pipelinebpred}
D.A. {Jimenez}, ``{Reconsidering complex branch predictors},'' The Ninth
  International Symposium on High-Performance Computer Architecture, 2003.
  HPCA-9 2003. Proceedings., pp.43--52, Feb\ 2003.

\bibitem{matsuibpred}
K.~{Matsui}, M.~{Ashraful Islam}, and K.~{Kise}, ``{An Efficient Implementation
  of a TAGE Branch Predictor for Soft Processors on FPGA},'' 2019 IEEE 13th
  International Symposium on Embedded Multicore/Many-core Systems-on-Chip
  (MCSoC), pp.108--115, Oct\ 2019.

\bibitem{nexys4ddr}
{Digilent, Inc.}, {Nexys 4 DDR Reference Manual}, rev.c~ed., 2016.

\bibitem{Dhrystone}
R.P. Weicker, ``{Dhrystone: A Synthetic Systems Programming Benchmark},''
  Commun. ACM, vol.27, no.10, pp.1013--1030, Oct.\ 1984.

\bibitem{coremark}
{EEMBC}, ``{CoreMark | CPU Benchmark – MCU Benchmark}.''
  https://www.eembc.org/coremark/.

\bibitem{riscvtests}
{RISC-V Foundation}, ``{riscv-tests}.'' https://github.com/riscv/riscv-tests.

\bibitem{riscvcoremark}
{UC Berkeley Architecture Research}, ``{Setup scripts and files needed to
  compile CoreMark on RISC-V}.'' https://github.com/riscv-boom/riscv-coremark.

\end{thebibliography}
%\nocite{*}

%\begin{thebibliography}{99}% more than 9 --> 99 / less than 10 --> 9
%\bibitem{}
%\end{thebibliography}

%%%%%%%%%%%%%%%%%%%%%%%%%%%%%%%%%%%%%%%%%%%%%%%%%%%%%%%%%%%%%%%%%%%%%%%%%%%%%%%%%%%%%%%%%%

%\profile{}{}
%\profile*{}{}% without picture of author's face

\profile[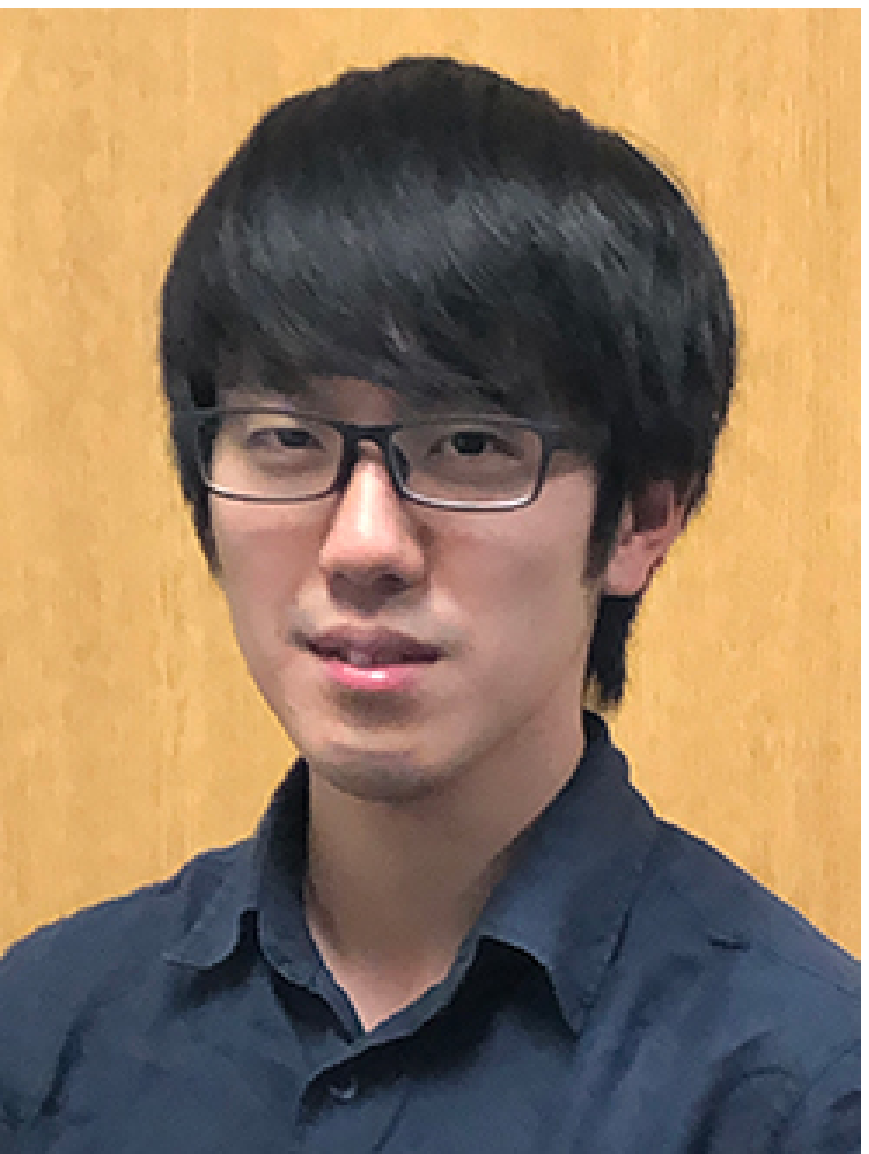]{Hiromu Miyazaki}{
received the B.E degrees in Department of Computer Science 
from Tokyo Institute of Technology, Japan in 2019.
He is currently a master course student of the Graduate School of Computing, 
Tokyo Institute of Technology, Japan.
His research interest is computer architecture and FPGA computing.
He is a student member of IEICE.
}
\profile[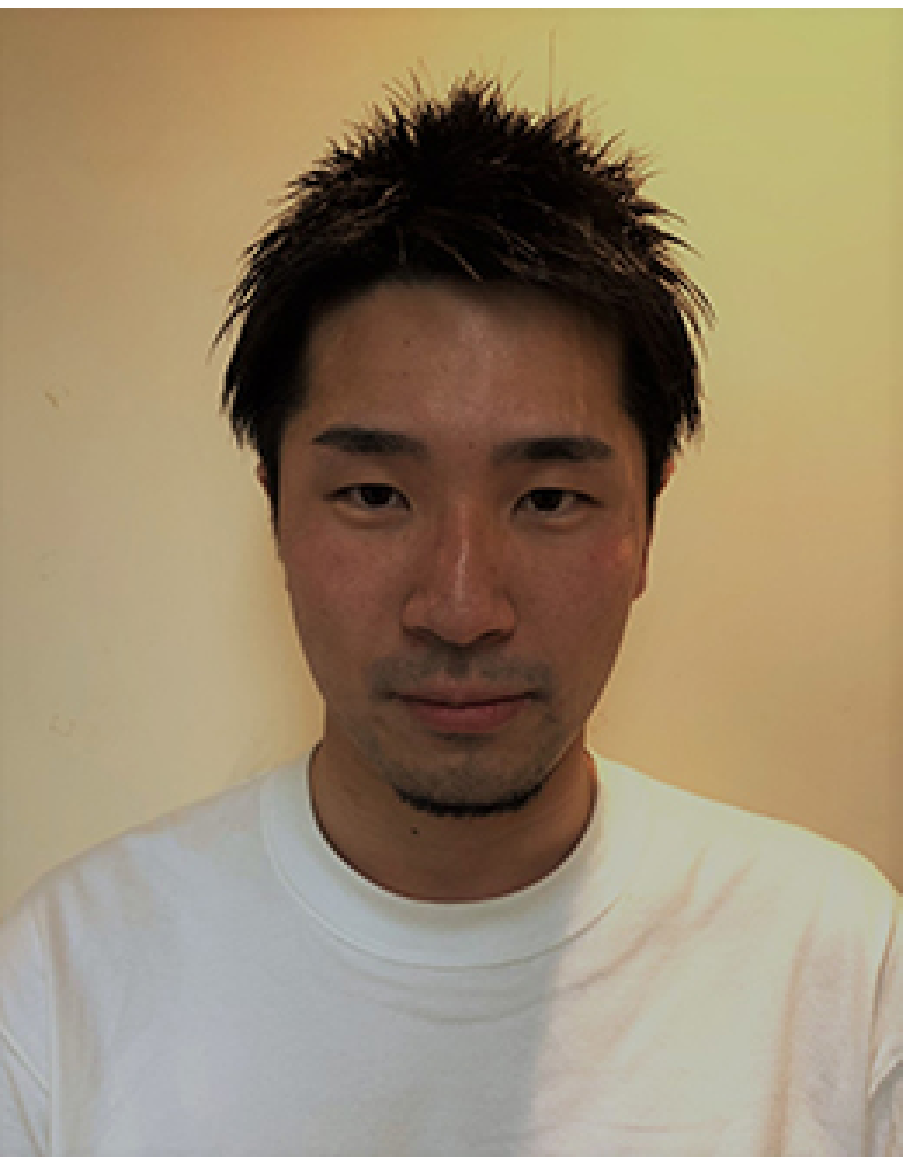]{Takuto Kanamori}{
is currently a bachelor course student of the School of Computing,
Tokyo Institute of Technology, Japan. 
His research interest is computer architecture and FPGA computing.
}
\profile[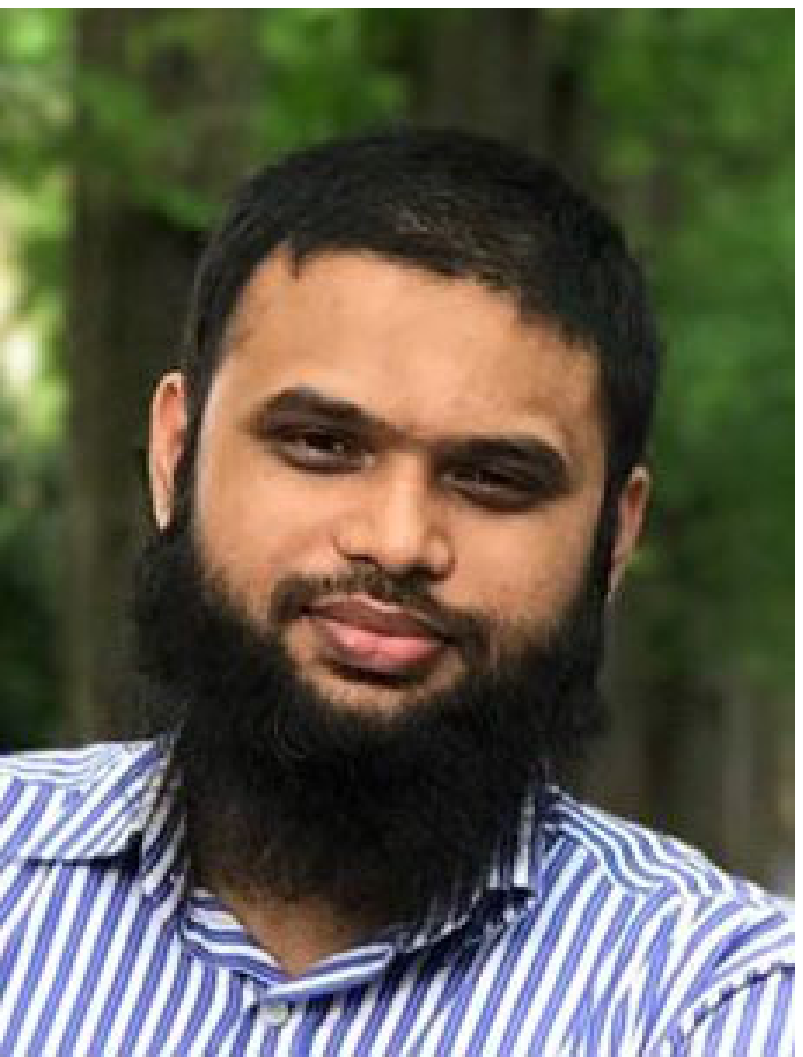]{Md Ashraful Islam}{
have graduated from the University of Rajshahi, Bangladesh.
I am a 1st-year doctoral student at the Tokyo Institute of Technology.
I have 8-years of experience in the Semiconductor Industry 
in ASIC, SoC design and Verification.
My research interest is in Computer Architecture, 
especially on Processor design and memory sub-system design.
}
\profile[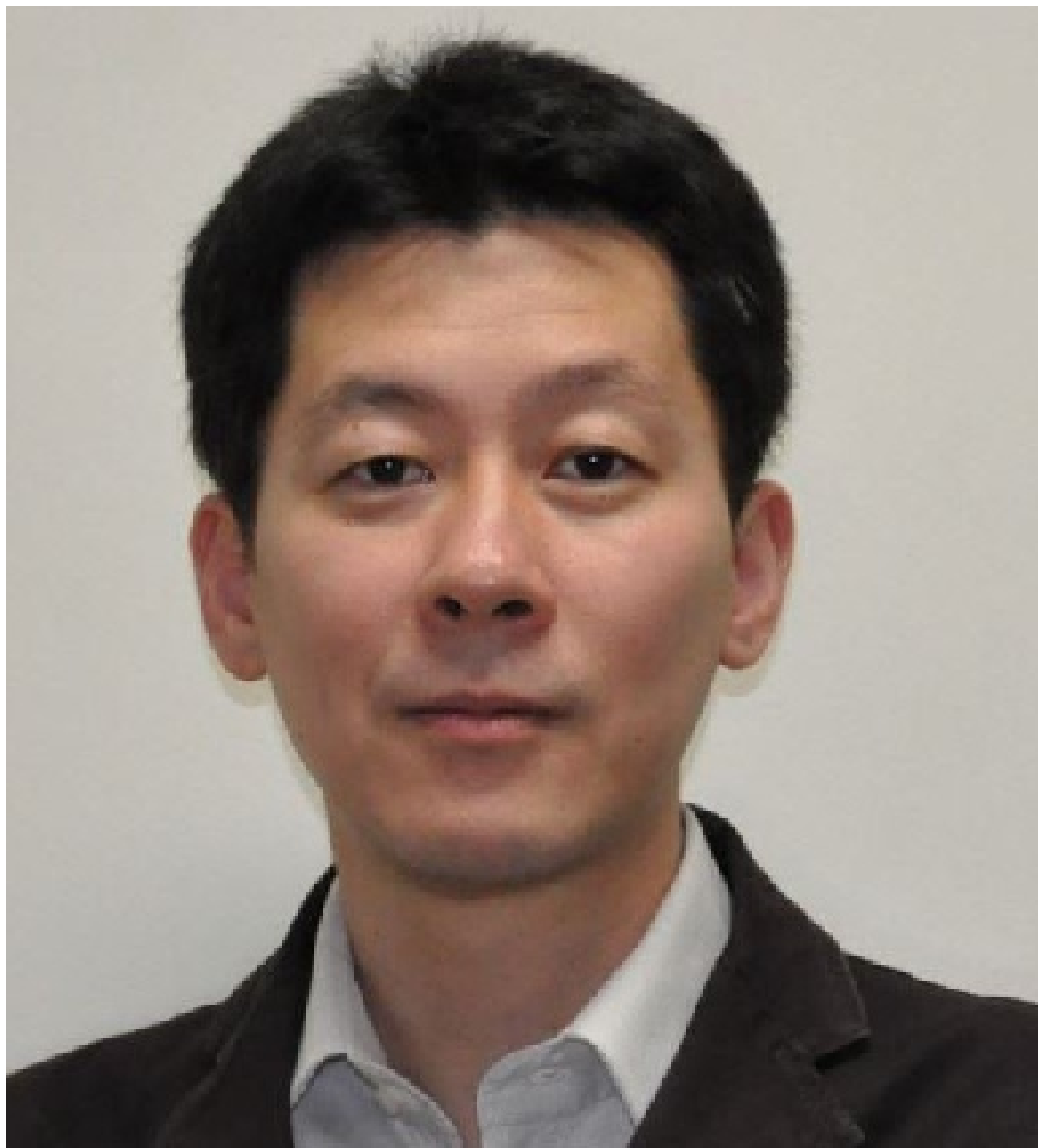]{Kenji Kise}{
received the B.E. degree from Nagoya University in 1995,
the M.E. degree and the Ph.D. degree
in information engineering from the University of Tokyo in 1997 and 2000, respectively. 
He is currently an associate professor of the School of Computing, 
Tokyo Institute of Technology. 
His research interests include computer architecture and parallel processing.
He is a member of ACM, IEEE, IEICE, and IPSJ.
}

%%%%%%%%%%%%%%%%%%%%%%%%%%%%%%%%%%%%%%%%%%%%%%%%%%%%%%%%%%%%%%%%%%%%%%%%%%%%%%%%%%%%%%%%%%

\end{document}